# Simultaneous Transitions in Cuprate Momentum-Space Topology and Electronic Symmetry Breaking

K. Fujita[1,2,3†], Chung Koo Kim[1†], Inhee Lee[1], Jinho Lee[1,4,5], M. H. Hamidian[1,2], I. A. Firmo[2], S. Mukhopadhyay[2,6], H. Eisaki[7], S. Uchida[3], M. J. Lawler[2,8], E. -A. Kim[2], and J. C. Davis[1,2,9,10]

[1] CMPMS Department, Brookhaven National Laboratory, Upton, NY 11973, USA.
[2] LASSP, Department of Physics, Cornell University, Ithaca, NY 14853, USA.
[3] Department of Physics, University of Tokyo, Bunkyo-ku, Tokyo 113-0033, Japan.
[4] Institute for Basic Science, Seoul 151-747, Republic of Korea
[5] Department of Physics and Astron., Seoul National University, Seoul 151-747, Korea.
[6] Cornell Center for Material Research, Cornell University, Ithaca, NY 14853, USA.
[7] Institute of Advanced Industrial Science and Technology, Tsukuba, Ibaraki 305-8568, Japan.
[8] Dept. of Physics and Astronomy, Binghamton University, Binghamton, NY 13902.
[9] School of Physics and Astronomy, University of St. Andrews, Fife KY16 9SS, Scotland.
[10] Kavli Institute at Cornell for Nanoscale Science, Cornell University, Ithaca, NY 14853, USA.

† These authors contributed equally to this project.

**The existence of electronic symmetry breaking in the underdoped cuprates, and its disappearance with increased hole-density $p$, are now widely reported. However, the relationship between this transition and the momentum-space ($\vec{k}$-space) electronic structure underpinning the superconductivity has not been established. Here we visualize the $\vec{Q}=0$ (intra-unit-cell) and $\vec{Q}\neq0$ (density wave) broken-symmetry states simultaneously with the coherent $\vec{k}$-space topology, for $Bi_2Sr_2CaCu_2O_{8+\delta}$ samples spanning the phase diagram $0.06 \leq p \leq 0.23$. We show that the electronic symmetry breaking tendencies weaken with increasing $p$ and disappear close to $p_c$=0.19. Concomitantly, the coherent $\vec{k}$-space topology undergoes an abrupt transition, from arcs to closed contours, at the same $p_c$. These data reveal that the $\vec{k}$-space topology transformation in cuprates is linked intimately with the disappearance of the electronic symmetry breaking at a concealed critical point.**

**2**     The highest known superconducting critical temperature $T_c$ (*1-3*) occurs atop the $T_c(p)$ 'dome' of hole-doped cuprates (Fig. 1A). In addition to the superconductivity, electronic broken-symmetry states (*4*) have also been reported at low $p$ in many such compounds. Wavevector $\vec{Q}$=0 (intra-unit-cell) symmetry breaking, typically of 90º-rotational ($C_4$) symmetry, is reported in $YBa_2Cu_3O_{6+\delta}$, $Bi_2Sr_2CaCu_2O_{8+\delta}$ and $HgBa_2CuO_{4+x}$ (*5-14*). Finite wavevector $\vec{Q}\neq0$ (density wave) modulations breaking translational symmetry, long detected in underdoped $La_{2-x-y}Nd_ySr_xCuO_4$ and $La_{2-x}Ba_xCuO_4$ (*15,16*), are now also reported in underdoped $YBa_2Cu_3O_{6+\delta}$, $Bi_2Sr_2CuO_{6+\delta}$ and $Bi_2Sr_2CaCuO_{8+\delta}$ (*17-25*). Summarizing all such reports in Fig. 1A reveals some stimulating observations. First, although the $\vec{Q}$=0 and $\vec{Q}\neq0$ states are detected by widely disparate techniques and are distinct in terms of symmetry, they seem to follow approximately the same phase-diagram trajectory (shaded band Fig. 1A) as if facets of a single phenomenon (*26*). The second implication is that a critical point (perhaps a quantum critical point) associated with these broken-symmetry states may be concealed beneath the $T_c(p)$ dome. Numerous earlier studies reported sudden alterations in many electronic/magnetic characteristics near $p$=0.19 (*2,3,27*), but whether these phenomena are caused by electronic symmetry changes (*28*) at a critical point was unknown.

**3**     In $\vec{k}$-space, the hole-doped cuprates also exhibit an unexplained transition in electronic structure with increasing hole density. Open contours or "Fermi arcs" (*29-32*) are reported at low $p$ in all compounds studied, while at high $p$ closed hole-like pockets surrounding $\vec{k} = (\pm1, \pm1)\pi/a_0$ are observed (*33,34*). One possibility is that such a transition could occur due to the disappearance of an electronic ordered state, with the resulting modifications to the Brillouin zone geometry altering the topology of the electronic bands (*28*).

**4**     Our strategy is therefore a simultaneous examination of both the $\vec{k}$-space electronic structure and the $\vec{Q}=0/\vec{Q}\neq 0$ broken-symmetry states, over a sufficiently wide range of *p* to include any concealed critical point. The objective is to search for a relationship between the broken symmetry states (*5-25*) and the Fermi surface topology (*29-34*). Fourier transform analysis of spectroscopic imaging scanning tunneling microscopy (SI-STM) is unique in allowing access simultaneously to the $\vec{Q}=0/\vec{Q}\neq 0$ broken-symmetry states (*26,35*) and to the $\vec{k}$-space structure by using quasiparticle scattering interference (QPI) (*35*). The SI-STM tip/sample differential tunneling conductance $g(\vec{r}, E = eV)$ at location $\vec{r}$ and energy $E$ relates to density-of-electronic-states $N(\vec{r}, E)$ as $g(\vec{r}, E) \propto \left[eI_s/\int_0^{eV_s} N(\vec{r}, E')dE'\right] N(\vec{r}, E)$ (*35*) ($I_s$ and $V_s$ are arbitrary parameters). In cuprates, it is necessary to study $Z(\vec{r}, |E|) = g(\vec{r}, E)/g(\vec{r}, -E)$ because it suppresses the severe systematic errors due to the unknown denominator $\int_0^{eV_s} N(\vec{r}, E')dE'$ by division, thereby allowing $\vec{q}$-vectors and symmetries to be measured correctly (*35*).

**5**   The $\vec{k}$-space structure of coherent Bogoliubov quasiparticles is then detectable because scattering between the eight joint density-of-states maxima at $\vec{k}_j(E); j = 1, 2,...,8$ (circles in inset Fig. 1B) produces interference patterns with wavevectors $\vec{q}(E) = \vec{k}_j(E) - \vec{k}_i(E)$ in $Z(\vec{q}, E)$, the Fourier transform of $Z(\vec{r}, E)$. At low *p*, one finds that the predicted set of seven inequivalent Bogoliubov QPI wavevectors $\vec{q}_m(E); m = 1,2,..7$ (inset Fig. 1B) exist only below an energy $\Delta_0$ (Fig. 1B), indicating that it is the limiting binding energy of a Cooper pair (*35*). At $|E|>\Delta_0$ the dispersive $\vec{q}_m$ disappear (Fig. 1B, Movie S1) and are replaced by broken-symmetry states consisting of: (i) spatial modulations with energetically quasistatic wavevectors $\vec{q}_1^*$ and $\vec{q}_5^*$ breaking translational symmetry (*24,25,35,36*) and (ii) $\vec{Q}=0$ (intra-unit-cell) breaking of $C_4$ symmetry detectable either directly in $\vec{r}$–space *(11,36)* or at the Bragg wavevectors (*11,26,35*). But the complete doping dependence of these broken-symmetry signatures was unknown.

**6**     To determine the $\vec{k}$-space topology of coherent states across the phase diagram of Bi$_2$Sr$_2$CaCu$_2$O$_{8+\delta}$, we use a recently developed approach (*37*) that requires measurement of only a single QPI wavevector: $\vec{q}_4$ (blue arrow in Fig. 1B, inset). Brillouin zone geometry (Fig. 2A) means that $\vec{q}_4$ is simply related to the Fermi surface (FS) states $\vec{k}_F$ as

$$\vec{q}_4\left(E = \Delta(\vec{k}_F)\right) = 2\vec{k}_F \qquad (1)$$

Here $\Delta(\vec{k})$ is the $d_{x^2-y^2}$ superconducting energy gap whose energy-minimum follows the trajectory of $\vec{k}_F$. An image of the locations of $\vec{q}_4(E)/2$ for all $0<|E|<\Delta_0$ (pale blue region in Fig. 1B) then yields the Fermi surface location of coherent states that contribute to Cooper pairing. An efficient way to locate these states is to sum all the $Z(\vec{q},E)$ images to the energy $\Delta_0$

$$\Lambda(\vec{q}) = \sum_{E\cong 0}^{\Delta_0} Z(\vec{q},E) \qquad (2)$$

and then to plot the contour of $\vec{q}_4$ within these $\Lambda(\vec{q})$ images [(*38*) Section I, Fig. S1, S2]. The power of this novel procedure is demonstrated in the determination of the FS in Fig. 2B [(*38*) Section II]. Applying this $\Lambda(\vec{q})$ approach to determine the doping dependence of $\vec{k}$-space topology, we find very different results at low and high *p*. Figure 2C shows $\Lambda(\vec{q})$ at *p*=0.14 while Fig. 2D shows $\Lambda(\vec{q})$ at *p*=0.23. The most prominent difference between the two is that the contour of $\vec{q}_4$ only spans four arcs in Fig. 2C whereas it completes four closed curves surrounding $\vec{q} = (\pm 1, \pm 1)2\pi/a_0$ in Fig. 2D. A recent preprint reports similar phenomena in Bi$_2$Sr$_2$CuO$_{6+d}$ (*37*). In Figure 2E we show the complete doping dependence of measured $\vec{q}_4/2$ over the full range of *p* [(38) Section II]. A striking transition in $\vec{k}$-space topology is observed within the narrow range *p*≈0.19±0.01 wherein the arc of coherent Bogoliubov states typical of low *p* suddenly switches to the complete closed contour surrounding $\vec{k}$=(±1, ±1)$\pi/a_0$ ((*38*) Section II, Fig. S3, Movie S2).

**7** Next we study the broken-symmetry states by examining $Z(\vec{r},E)$ measured simultaneously with the $\vec{k}$-space data in Fig. 2, but now for $\Delta_0<|E|<\Delta_1$ (pink regions in Fig. 1B), where $\Delta_1$ is the maximum detectable gap (pseudogap at low $p$ and maximum superconducting gap at high $p$ (*35*)). These images exhibit several distinct broken spatial symmetries whose evolution with $p$ we explore. Figure 3A shows $Z(\vec{r},E\sim\Delta_1)$ for $p$=0.08 while Fig. 3B shows $Z(\vec{r},E\sim\Delta_1)$ for $p$=0.23, with their Fourier transforms $Z(\vec{q},E\sim\Delta_1)$ shown in Fig. 3C and 3D respectively. The former exhibits the widely reported (*24,25,26,35,36*) quasistatic wavevectors $\vec{q}_1^*$ and $\vec{q}_5^*$ of states with local symmetry breaking along with the Bragg peaks (red circle) while, in the latter, the quasistatic wavevectors $\vec{q}_1^*$ and $\vec{q}_5^*$ have disappeared. The $\vec{Q}$=0 broken $C_4$-symmetry states can be detected by using the lattice-phase-resolved nematic order parameter (*11*)

$$O_N^q(E) = ReZ(\vec{Q}_y,E) - ReZ(\vec{Q}_x,E) \qquad (3)$$

The $\vec{Q}_x$ and $\vec{Q}_y$ are the Bragg vectors after the necessary transformation to perfect lattice periodicity in $Z(\vec{r},E)$ so that real and imaginary components of the Bragg amplitudes, $ReZ(\vec{Q},E)$ and $ImZ(\vec{Q},E)$, are well defined (*11,35*). The measured $O_N^q(\vec{r},E\sim\Delta_1)$ for $p$=0.06 and $O_N^q(\vec{r},E\sim\Delta_1)$ for $p$=0.23 are shown in Figs. 3E, F respectively [(*38*) Section IV]. Here we see that the extensive order in $O_N^q(\vec{r},E\sim\Delta_1)$ observed at low $p$ (*11*) has disappeared at high $p$, leaving nanoscale domains (*26*) probably nucleated by disorder. The doping dependence of $I(\vec{q}_5^*)$, the intensity of the $\vec{q}_5^*$ modulations in $Z(\vec{q},E)$, is shown in Fig. 3G [(*38*) Section III Figs. S4, S5], while the dependence of the spatially averaged magnitude $|<O_N^q(\vec{r},E\sim\Delta_1)>|$ of the $\vec{Q}$=0 $C_4$ breaking is shown in Fig. 3H [(*38*) Section IV Fig. S7, S8]. These plots reveal that the more extended $\vec{Q}$=0 broken symmetry and the shorter-range ordering tendencies in $\vec{Q}\neq0$ modulations (*11,26,35*) disappear near a critical doping $p_c\approx0.19$.

*8*     Figure 4A is a schematic summary of our findings, from Bogoliubov QPI techniques (*32,35,37*), on the dependence of $\vec{k}$-space electronic structure with increasing *p* Figure 4B shows that the wavevectors $\vec{k} = \vec{q}_E/2$ of states at which Bogoliubov QPI disappears (closed circles) evolve along the $\vec{k}$-space lines $(\pm 1, 0)\pi/a_0 \rightarrow (0, \pm 1)\pi/a_0$ with increasing *p*. Concomitantly, the quasistatic wavevectors $\vec{q}_1^*/2$ and $(2\pi - \vec{q}_5^*)/2$ of broken-symmetry states also evolve on the same trajectory (closed squares). Thus the $\vec{q}_1^*$ and $\vec{q}_5^*$ wavevectors of incommensurate (density wave) modulations evolve with doping as shown in Fig. 4C (*35*). Figure 4D shows the area of $\vec{k}$-space between the arc and the line $(1,0)\pi/a_0 \rightarrow (0,1)\pi/a_0$ (left inset) increasing proportional to hole-density *p (32,35)*; at *p*=0.19 there is a transition to a diminishing area of electron count as 1-*p* for the closed-contour FS topology. Finally, we show in Figure 4E-G that the critical point $p_c \approx 0.19$ is associated microscopically with a transition to conventional *d*-wave Bogoliubov QPI on a complete Fermi surface (simulated in 4E and measured at *p*>$p_c$ in 4G) from a highly distinct form of scattering (Fig. 4F) of unknown cause (*39*).

*9*     To recapitulate: with increasing hole-density the $\vec{Q} \neq 0$ modulations (density waves) weaken and disappear at $p_c \approx 0.19$ (Figs. 1A, 3A-D, 3G). Concurrently, the $\vec{Q}=0$ broken-symmetry (intra-unit-cell nematic) states become progressively more disordered (*13*) and reach a zero average value at approximately the same *$p_c$* (Figs. 1A, 3E-F, 3H). Simultaneously, the $\vec{k}$-space topology of coherent Bogoliubov quasiparticles (or the Fermi surface supporting their superconducting gap) undergoes an abrupt transition from 'arcs' to closed contours (Figs 2,4, (38) Movie S2). This key transformation of cuprate electronic structure is therefore linked directly with the disappearance of the electronic symmetry breaking. However, this phenomenology also exhibits many peculiar components unexpected within a simple Fermi surface reconstruction scenario. First, the co-evolution and contiguous disappearance at $p_c$ of the signatures of two distinct broken symmetries (Figs 1A, 3), reinforces the deductions that they are microscopically closely related (*11,26,35,36,40*). Second, because the $\vec{Q}{\neq}0$ modulations exhibit wavevectors generated by scattering regions ('hot spots') moving along the $\vec{k}$-space lines $(\pm1,0)\pi/a_0 \rightarrow (0,\pm1)\pi/a_0$ (Figs 4A,B,C; (*32,35)*), Fermi surface nesting provides an inadequate explanation for the cuprate density waves. Third, the abrupt $\vec{k}$-space topology change at $p_c$ (Figs 2E and 4A,F,G) exhibits characteristics more reminiscent of an antinodal 'coherence recovery' transition (*41*) than of a conventional band reorganization. Fourth, because the disappearance of the pseudogap is associated axiomatically with the reappearance of coherent antinodal states, and because the latter is precisely what occurs at $p_c$ (Fig. 2E, 4A), the pseudogap (*1-3*) and the electronic symmetry breaking (*5-25*) must be intimately linked (Fig. 1A). Finally, as neither long-range $\vec{Q}{\neq}0$ order nor any associated quantum critical point can exist with quenched disorder (*40*), a nematic critical point at which the electronic symmetry breaking between the two oxygen sites within the CuO$_2$ unit cell (*11,26,35,36*) disappears, seems most consistent with our observations.

# Figure Captions

### Figure 1 Hole–density dependence of cuprate broken-symmetry states

A. Phase diagram of hole-doped cuprates showing the $T_c(p)$ "dome" (blue line). Symbols: The onset temperature of two types of symmetry breaking in $Bi_2Sr_2CuO_{6+\delta}$($B_1$), $Bi_2Sr_2CaCu_2O_{8+\delta}$($B_2$), $YBa_2Cu_3O_{6+\delta}$(Y), and $HgBa_2CuO_{4+x}$(H). Circular symbols: $\vec{Q}=0$ (intra-unit-cell) symmetry breaking. Diamond symbols: $\vec{Q}\neq 0$ broken translational symmetry ("density waves").

B. Dashed black line: typical differential conductance spectrum $g(E)$ of underdoped $Bi_2Sr_2CaCu_2O_{8+\delta}$ here at $p$=0.06. Pale blue shaded region indicates where the dispersive $\vec{q}_m(E); m = 1,2,..7$ Bogoliubov QPI wavevectors indicative of Cooper pair breaking exist (color coded in inset); they disappear at energy $\Delta_0$. Solid curves: simulation for QPI in a $d_{x^2-y^2}$ symmetry superconductor. Pink shaded regions exhibit quasistatic conductance modulations ('density waves') exhibiting wavevectors $\vec{q}_1^*$ and $\vec{q}_5^*$.

### Figure 2 Momentum-space topology transition from Bogoliubov QPI

A. Schematic of d-wave superconducting energy gap $\Delta(\vec{\Box}_\Box)$ on two opposing segments of a Fermi surface; $\vec{q}_4\left(E = \Delta(\vec{k}_F)\right) = 2\vec{k}_F$ is indicated for several $\Delta(\vec{k}_F)$ by colored arrows.

B. The location of the Fermi surface is identified from the measured wavevectors $\vec{k} = \vec{q}_4(0 < E < \Delta_0)/2$. Shown are the resulting $\vec{k}_F$ for a sample with $p$=0.23; $\Delta_0 = \Delta_1 = 20 \pm 1 meV$

C. The measured $\Lambda(\vec{q})$ for $p$=0.14 sample; $\Delta_0 = 34 \pm 1 meV$. No complete contour for $\vec{q}_4$ can be detected; instead the coherent Bogoliubov quasiparticles are restricted to four arcs terminating at lines joining $(\pm 1,0)\pi/a_0$ to $(0,\pm 1)\pi/a_0$.

D. The measured $\Lambda(\vec{q})$ for $p$=0.23 sample; $\Delta_0 = 20 \pm 1 meV$. A complete closed contour for $\vec{q}_4$ surrounding $(\pm 1, \pm 1)\pi/a_0$ can be identified immediately.

E. The measured doping dependence of the $\vec{k}$-space topology of coherent Bogoliubov quasiparticles using the $\vec{q}_4$ technique. The transition from arcs terminating at the lines $(\pm1,0)\pi/a_0$ to $(0,\pm1)\pi/a_0$ to complete hole-pockets surrounding $(\pm1,\pm1)\pi/a_0$ at $p\approx0.19$ is evident.

**Figure 3 Measurements of hole–density dependence of $\vec{Q}=0$ and $\vec{Q}\neq0$ ordering**

A. $Z(\vec{r}, E\sim\Delta_1)$ for $p=0.08$. Incommensurate conductance modulations are clearly seen; $\Delta_1 = 84 \pm 1 meV$

B. $Z(\vec{r}, E\sim\Delta_1)$ for $p=0.23$ seen; $\Delta_1 = 20 \pm 1 meV$. No specific $q$-vector for modulations is seen, although the QPI signature of Bogoliubov quasiparticles does produce a jumbled standing wave pattern.

C. $Z(\vec{q}, E\sim\Delta_1)$ for $p=0.08$ from 3A; $\vec{q}_1^*$ and $\vec{q}_5^*$ wavevectors are indicated using violet and orange circles, respectively. Bragg peaks are shown using red circles.

D. $Z(\vec{q}, E\sim\Delta_1)$ for $p=0.23$ from 3B. No specific broken-symmetry state wavevectors are apparent, while the residual dispersive effects of Bogoliubov quasiparticles are seen.

E. $\vec{Q}=0$ C$_4$ broken-symmetry order parameter $O_N^q(\vec{r}, E\sim\Delta_1)$ for $p=0.06$; seen; $\Delta_1 = 124 \pm 1 meV$ This whole field of view is a single color indicating that long range $\vec{Q}=0$ intra-unit-cell C$_4$ symmetry breaking exists.

F. Intra-unit-cell broken C$_4$ symmetry $O_N^q(\vec{r}, E\sim\Delta_1)$ for $p=0.22$; $\Delta_1 = 28 \pm 1 meV$; long-range order has been lost but nanoscale domains of opposite nematicity persist. Broken circle represents the spatial resolution of the analysis.

G. Intensity of incommensurate modulations with wavevector $\vec{q}_5^*$, $I(\vec{q}_5^*)$, initially increases upon doping peaking near $p\sim1/8$, and then diminishes to reach zero at $p\approx0.19$.

H. Filled square symbols indicate measured spatial average value of the $\vec{Q}$=0 broken C4 symmetry $|<O_N^q(\vec{r},E\sim\Delta_1)>|$ which diminishes steadily with increasing p, to reach zero near p=0.19. Triangles indicate measured standard deviation of $O_N^q$ : $\delta O_N^q(\vec{r},E\sim\Delta_1)$ .

**Figure 4 Interlinked k-space structure of Bogoliubov QPI and $\vec{Q}$≠0 symmetry breaking at lines joining $(\pm 1, 0)\pi/a_0$ to $(0, \pm 1)\pi/a_0$**

A. Schematic of $\vec{k}$-space locus of states generating Bogoliubov QPI with increasing hole density in $Bi_2Sr_2CaCu_2O_{8+\delta}$. An abrupt transition occurs at p≈0.19.

B. The measured wavevectors $\vec{k}$ of states at which Bogoliubov QPI disappears $\vec{q}_E/2$ (closed circles), and those of the quasistatic broken-symmetry modulations with $\vec{q}_1^*/2$ and $(2\pi - \vec{q}_5^*)/2$ (closed squares).

C. The measured doping dependence of wavevectors of incommensurate conductance modulations ('density waves') $\vec{q}_1^*$ and $\vec{q}_5^*$ derived from B (*35*).

D. The $\vec{k}$-space area between the arc and the lines joining $(\pm 1,0)\pi/a_0$ to $(0,\pm 1)\pi/a_0$ is proportional to p. With the appearance of the closed FS at p≈0.19, there is a transition to a diminishing area of electron count 1-p.

E. T-matrix scattering interference simulation for $\Lambda(\vec{q})$ for a complete Fermi surface and *d*-wave gap with conventional (time-reversal preserving) scattering.

F. Measured $\Lambda(\vec{q})$ for p=0.14 ( $\Delta_0 = 34 \pm 1meV$ ) is in sharp contrast to the simulation result in E.

G. For p=0.23 $\Lambda(\vec{q})$ conforms closely to the conventional d-wave Bogoliubov scattering scheme as anticipated in E.

**Acknowledgements**: We are particularly grateful to S. Billinge, J. E. Hoffman, S. A. Kivelson, D.-H. Lee and A. P. Mackenzie for key scientific advice. We thank K. Efetov, E. Fradkin, P. D. Johnson, J. W. Orenstein, C. Pepin, S. Sachdev and K. M. Shen for helpful discussions and communications. Experimental studies were supported by the Center for Emergent Superconductivity, an Energy Frontier Research Center, headquartered at Brookhaven National Laboratory and funded by the U.S. Department of Energy under DE-2009-BNL-PM015, as well as by a Grant-in-Aid for Scientific Research from the Ministry of Science and Education (Japan) and the Global Centers of Excellence Program for Japan Society for the Promotion of Science. C. K. K. acknowledges support from the FlucTeam program at Brookhaven National Laboratory under contract DE-AC02-98CH10886. JL acknowledges support from the Institute for Basic Science, Korea. IAF acknowledges support from Fundação para a Ciência e a Tecnologia, Portugal under fellowship number SFRH/BD/60952/2009. S.M. acknowledges support from NSF Grant DMR-1120296 to the Cornell Center for Materials Research. Theoretical studies at Cornell University were supported by NSF Grant DMR-1120296 to Cornell Center for Materials Research and by NSF Grant DMR-0955822. The original data is archived by Davis Group BNL & Cornell.

Figure 1

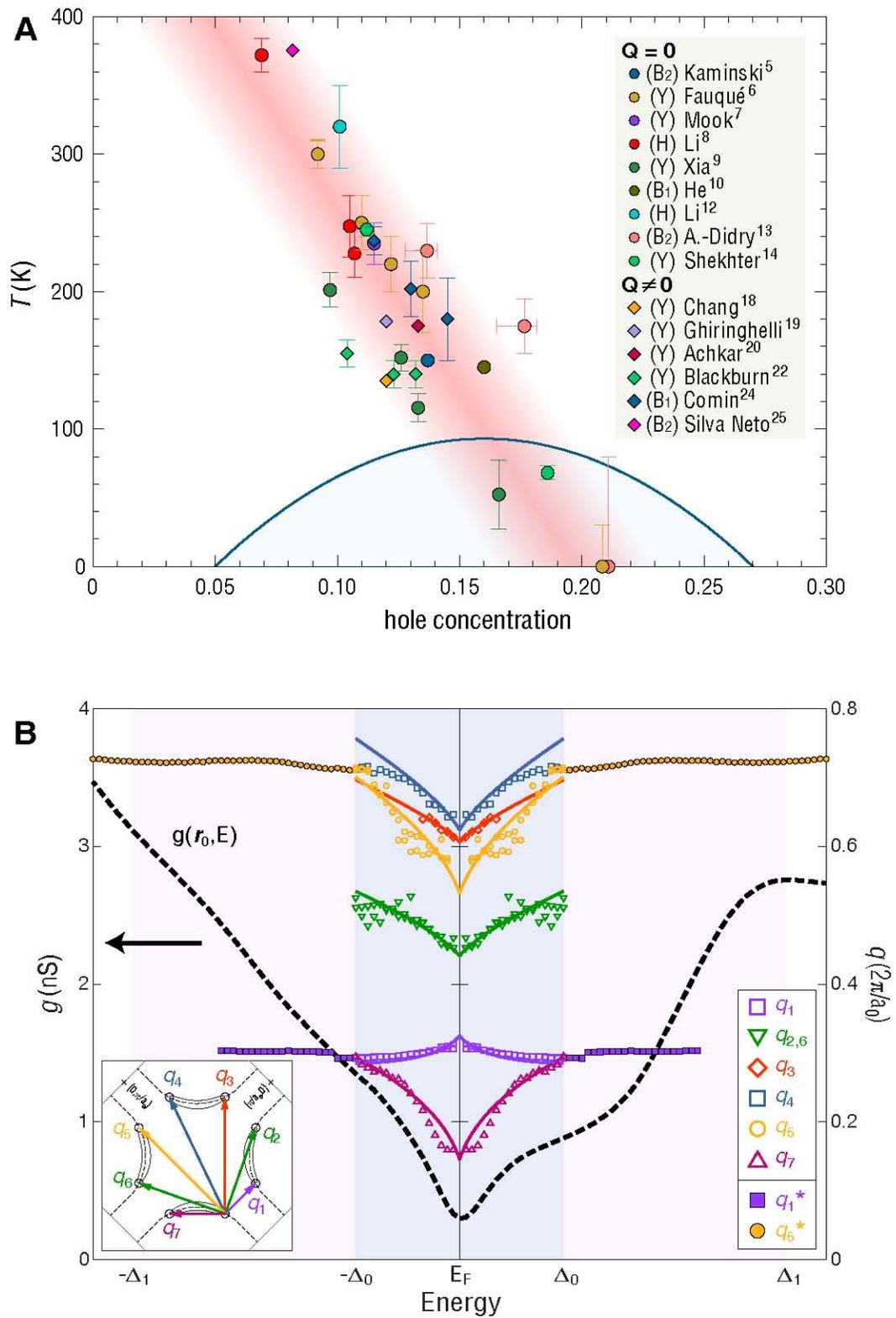

Figure 2

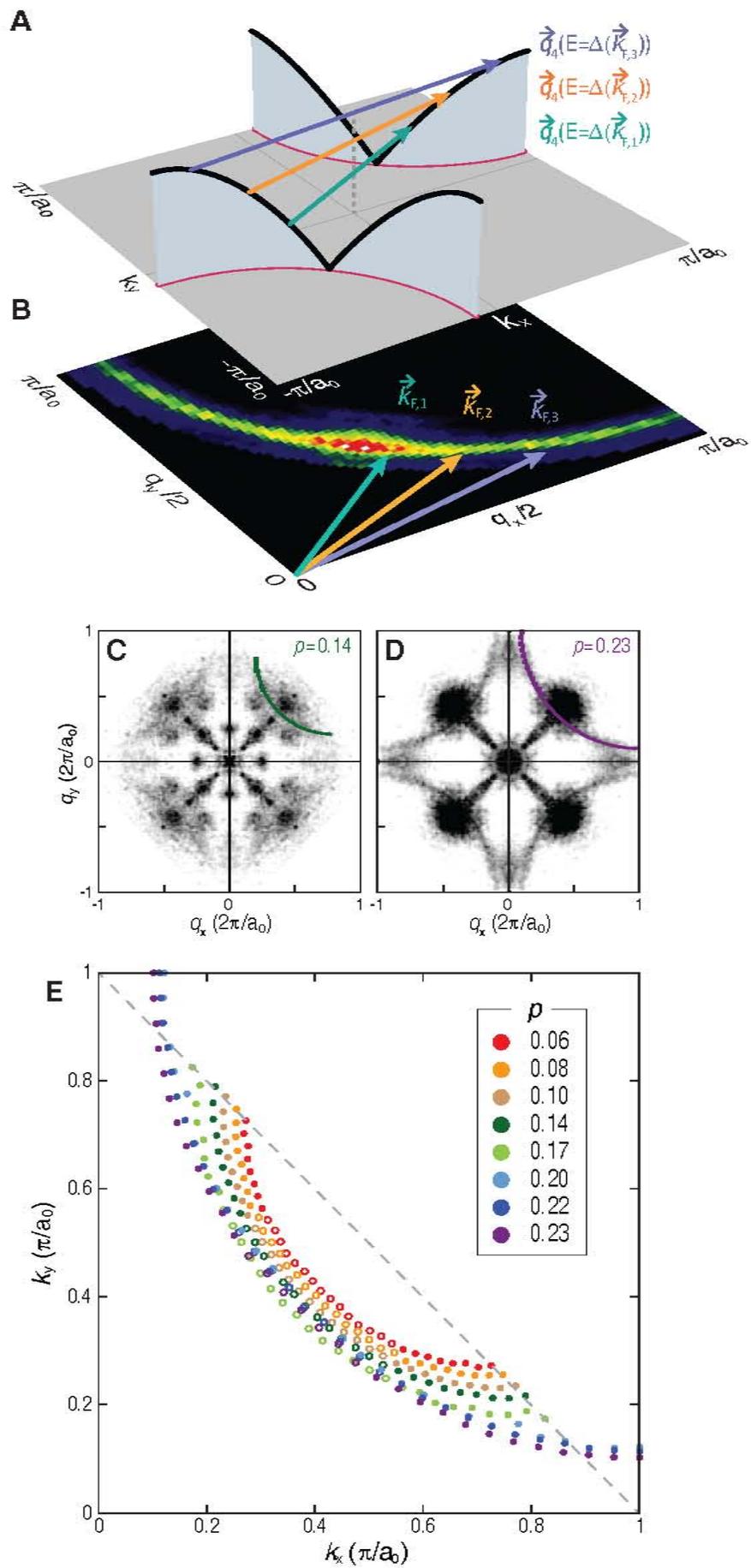

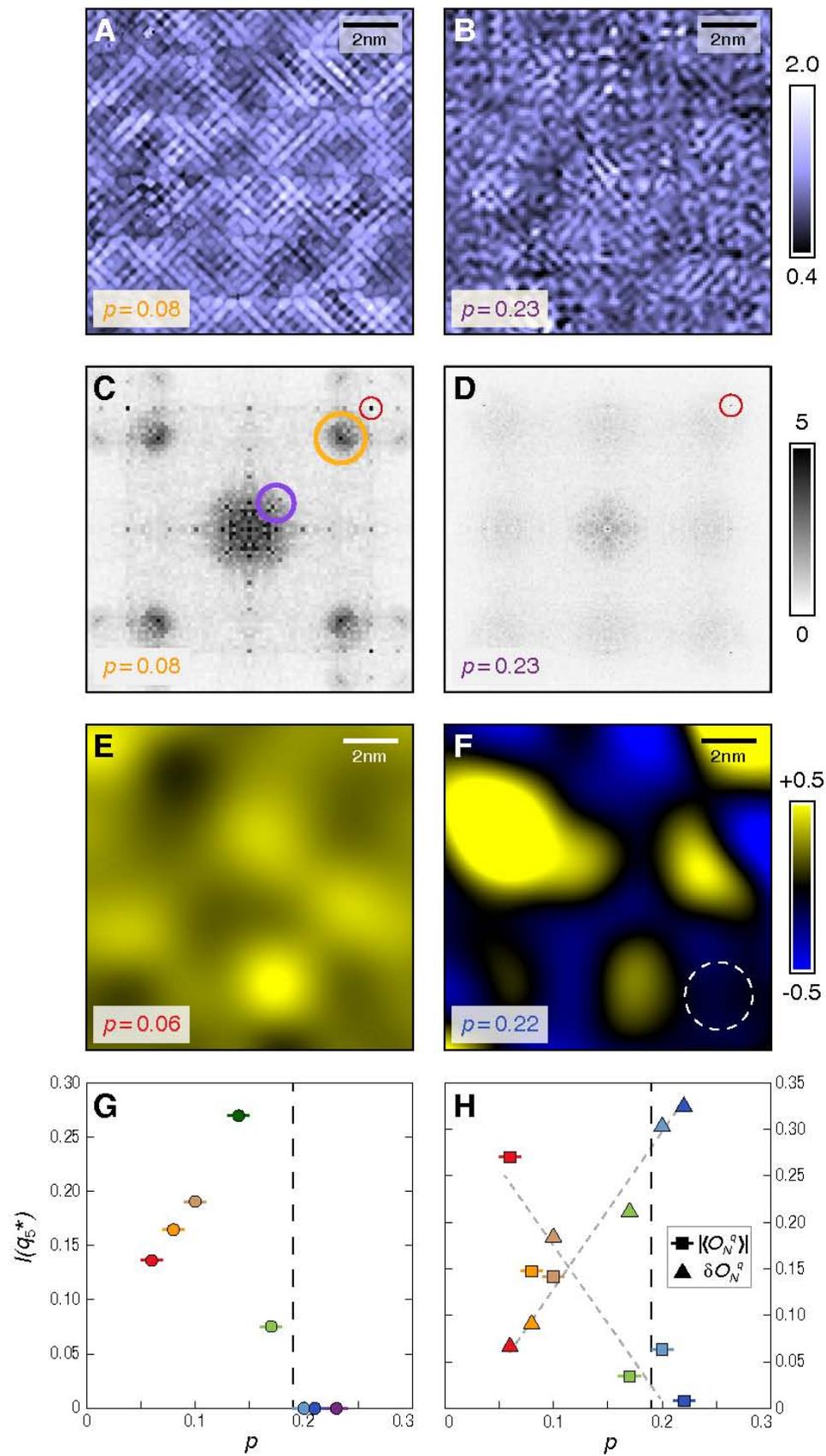

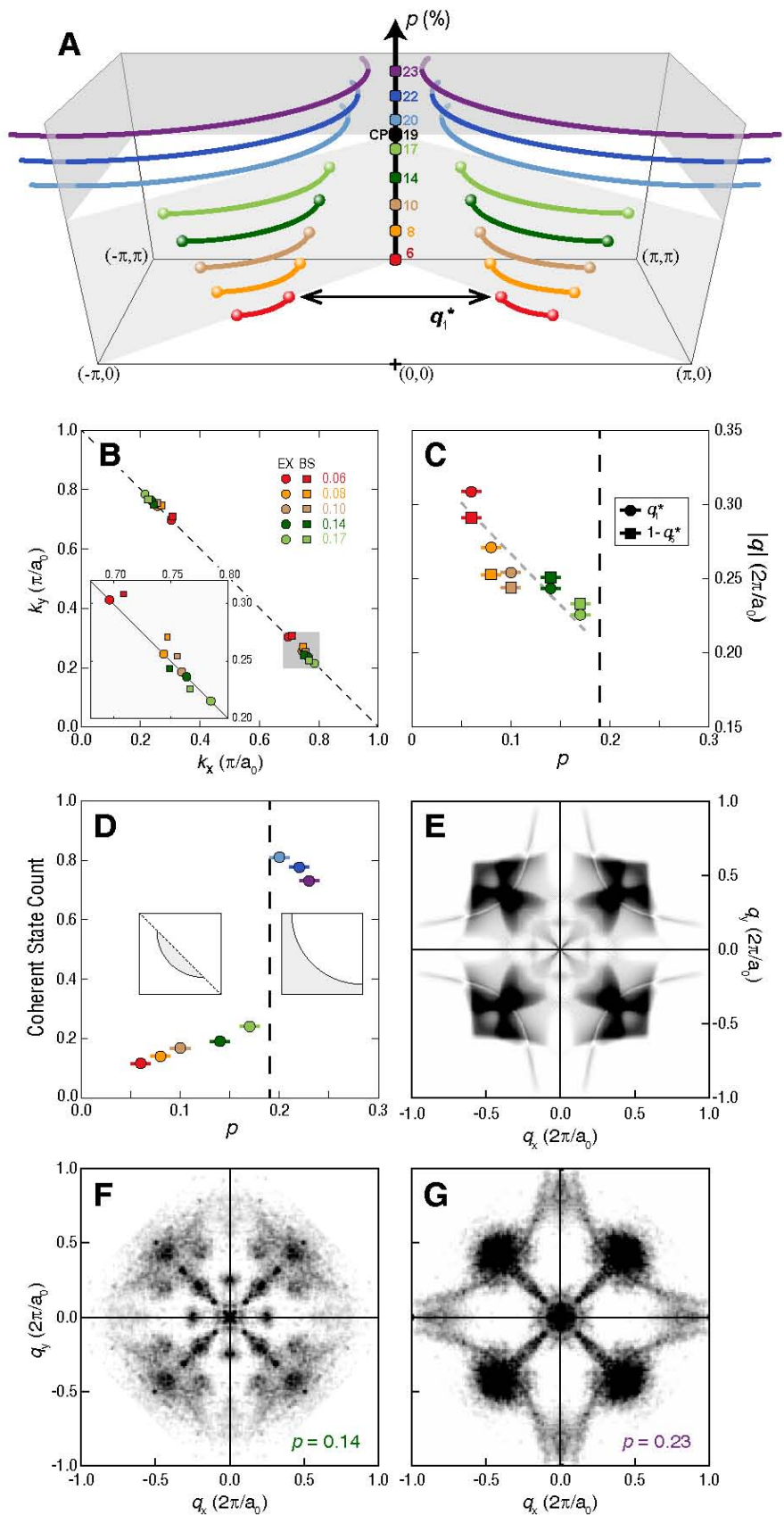

# Supplementary Materials for

## Simultaneous Transitions in Cuprate Momentum-Space Topology and Electronic Symmetry Breaking

K. Fujita[†], C. K. Kim[†], Inhee Lee, Jinho Lee, M. H. Hamidian, I. Firmo,

S. Mukhopadhyay, H. Eisaki, S. Uchida, M. J. Lawler, E. –A. Kim, and J. C. Davis*

*To whom correspondence should be addressed. E-mail: jcseamusdavis@gmail.com

**This PDF file includes:**

    Materials and Methods

    Supplementary Text

    Figs. S1 to S9

    Captions for Movies S1 to S2

**Other Supplementary Materials for this manuscript includes the following:**

    Movies S1 to S2

## Materials & Methods

High-quality single crystals of Bi$_2$Sr$_2$CaCu$_2$O$_8$ were synthesized for our studies using the traveling-solvent-floating-zone (TSFZ) method. The samples are of Bi$_{2.1}$Sr$_{1.9}$CaCu$_2$O$_{8+\delta}$ and Bi$_{2.2}$Sr$_{1.8}$Ca$_{0.8}$Dy$_{0.2}$Cu$_2$O$_{8+\delta}$ and were synthesized from well-dried powders of Bi$_2$O$_3$, SrCO$_3$, CaCO$_3$, Dy$_2$O$_3$, and CuO. The crystal growth was carried out in air and at growth speeds of 0.15 – 0.2 mm/h for all the samples. Inductively coupled plasma (ICP) spectroscopy was used for the composition analysis and a superconducting quantum interference device (SQUID) magnetometer was used for measurement of $T_c$. $T_c$ is defined as the onset temperature at which the zero-field-cooled susceptibility starts to drop. Oxidation annealing is performed in air or under oxygen gas flow, and deoxidation annealing is done in vacuum or under nitrogen gas flow for the systematic study at different hole-densities

We have studied sequence of Bi$_2$Sr$_2$CaCu$_2$O$_{8+\delta}$ samples for which $p \cong$ 0.23, 0.22, 0.20, 0.17, 0.14, 0.10, 0.08, 0.06. Each sample is inserted into the cryogenic ultra high vacuum of the SI-STM system, cleaved to reveal an atomically clean BiO surface, and all measurements were made between 1.9 K and 5 K. Three cryogenic SI-STM's were used during these studies. The basic spectroscopic imaging STM consists of lock-in-amplifier measurement of the differential tunneling conductance $dI/dV(\vec{r}, E = eV) \equiv g(\vec{r}, E = eV)$ with atomic resolution and register and as a function of both location **r** and electron energy $E$.

## Supporting Text and Figures

**(I) How normal state Fermi surface can be determined by using the $\vec{q}_4(E = \Delta_k)$**

Here we simulate the interference pattern created by the scattering of Bogoliubov quasiparticles within a framework of the *T*-matrix approximation for a simple *d*-wave superconductor with point-like impurities; we show that the location of the coherent Fermi surface $k_F$ can be visualized by using the interference patterns generated by quasiparticle scattering – specifically via $\vec{q}_4$. We start with a model Hamiltonian given by

$$H = \sum_{k\sigma} \varepsilon_k c^+_{k\sigma} c_{k\sigma} - \sum_k \left( \Delta_k c_{k\uparrow} c_{-k\downarrow} + h.c. \right) \tag{S1}$$

In the above, tight binding band $\varepsilon_k$ is given by [42]

$$\begin{aligned}\varepsilon_k = &\mu \\ &+ \frac{t}{2}(\cos k_x + \cos k_y) \\ &+ t'\cos k_x \cos k_y \\ &+ \frac{t''}{2}(\cos 2k_x + \cos 2k_y) \\ &+ \frac{t'''}{2}(\cos 2k_x \cos k_y + \cos k_x \cos 2k_y) \\ &+ t''''\cos 2k_x \cos 2k_y\end{aligned} \quad (S2)$$

where chemical potential $\mu=0.1305$(eV) and hopping parameters $t = -0.5908, t' = 0.0962, t'' = -0.1306, t''' = -0.0507, t'''' = 0.0939$(eV). Superconducting gap takes the form of $\Delta_k = \frac{\Delta_0}{2}(\cos k_x - \cos k_y)$, where $\Delta_0=0.030$(eV). Bare single particle Green's function in the superconducting state (2x2) is given by

$$G_0^{-1}(\vec{k},\omega) = (\omega + i\delta)I - \varepsilon_k \sigma_3 - \Delta_k \sigma_1 \quad (S3)$$

where σ's are the Pauli matrices. In Fig. S1, spectral function $(= -(1/\pi)\text{Im}G_0(\vec{k},\omega))$ at different energies are shown to exhibit the energy evolution of the momentum space electronic structure of the d-wave superconductor. T-matrix (2x2) that describes elastic scattering off a single point-like impurity is given by

$$T^{-1}(\omega) = (V_s\sigma_3 + V_m I)^{-1} - \int \frac{d^2k}{(2\pi)^2} G_0(\vec{k},\omega) \quad (S4)$$

Total density of states $n(\vec{r},\omega)$ now becomes $n(\vec{r},\omega) = n_0(\omega) + \delta n(\vec{r},\omega)$ where $n_0(\omega)$ is the density states without impurity and $\delta n(\vec{r},\omega)$ is the density of states generated by impurity scattering which is calculated by

$$\begin{aligned}A(\vec{r},\omega) &\equiv G_0(\vec{r},\omega)T(\omega)G_0(-\vec{r},\omega) \\ \delta n(\vec{r},\omega) &= -\frac{1}{\pi}\text{Im}(A_{11}(\vec{r},\omega) + A_{22}(\vec{r},-\omega))\end{aligned} \quad (S5)$$

Measurements of $\delta n$ in Eqn. S5 are complicated by the "setup effect" - a critical source of systematic error in SI-STM studies. It occurs because at very low temperature the STM tip-sample tunneling current is given by

$$I(\vec{r}, z, V) = f(\vec{r}, z) \int_0^{eV} n(\vec{r}, E) dE \tag{S6}$$

where $z$ is the tip-surface distance, $V$ the tip-sample bias voltage, $n(\vec{r}, E)$ the sample's local density of electronic states, while $f(\vec{r}, z)$ contains both effects of tip elevation and of spatially-dependent tunneling matrix elements. The $g(\vec{r}, E)$ data are then related to $n(\vec{r}, E)$ by,

$$g(\vec{r}, E = eV) = \frac{eI_S}{\int_0^{eV_S} n(\vec{r}, E') dE'} n(\vec{r}, E) \tag{S7}$$

where $V_S$ and $I_S$ are the junction stabilization bias voltage and current respectively. Thus $g(\vec{r}, E = eV)$ cannot be used to measure an unknown $n(\vec{r}, E)$, nor can $g(\vec{q}, E)$, the Fourier transform of $g(\vec{r}, E)$, ever be used to correctly measure the energy dependence of the wavevectors of density modulations.

Practically, in order to avoid the 'setup effect', we take ratio of the total density of states between the same energies but opposite sign, $Z(\vec{r}, \omega) \equiv \frac{g(\vec{r}, \omega)}{g(\vec{r}, -\omega)}$ and take the power spectral density Fourier transform to get $Z(\vec{q}, \omega)$. Then, finally, $\Lambda(\vec{q}, E) = \sum_{\omega \cong 0}^{E} Z(\vec{q}, \omega)$ is calculated to, in an efficient way, locate the states followed by $\vec{q}_4$ trajectory so as to visualize the coherent Fermi surface. As shown in Fig. S1 A2-E2, the $\vec{q}_4$ trajectory when plotted at all energies clearly follows the normal state Fermi surface. When summed up to the superconducting gap maximum (Fig. S1, E2) measurements of $\vec{q}_4$ therefore reproduce the $k$-space $E$=0 electronic structure of the non-superconducting state, i.e. the Fermi surface.

Following the procedure described above and using the Eqn. (2) in the main text, the experimentally observed $Z(\vec{q}, E)$ images are summed up to $\Delta_0$ to locate the states traced by $\vec{q}_4$, where $\Delta_0$ is the energy at which the "octet" signature of 7 interference wavevectors with internally consistent dispersions signifying the Bogoliubov quasiparticles, vanishes[32, 35]. The measured doping dependence of $\Lambda(\vec{q})$ is shown in Fig. S2 and from it the evolution of the states located using this $\vec{q}_4$ analysis is determined. These states are extracted and represented by dots for all the dopings in the Fig. S2. Fitting to these points by using a quarter circle, a good approximated normal state Fermi surfaces is then represented by solid curves. One can easily notice that below $p \approx 0.19$, the arc of coherent Bogoliubon is always observed to be terminated at the lines from $\vec{k} = \pm(\pi/a_0, 0)$ to $\pm(0, \pi/a_0)$. Here we see directly that, with increasing doping, the coherent *k*-space topology undergoes an abrupt change at $p\approx0.19$ from four arcs to four closed hole-like contours centered at $\vec{k} = \pm\ (\pi/a_0,\pi/a_0)$.

### (II) Normal state Fermi surface obtained by filtered $\vec{q}_4$ trajectory

Suppression of unwanted contributions from other *q*-vectors in $\Lambda(\vec{q},E)$ greatly helps to better visualize the doping evolution of the measured $\vec{q}_4$ trajectory. Here we describe how to emphasize the $\vec{q}_4$ trajectory only, with an application of filter function $W(R)$. First, the weighting function $W(R)$ is calculated by integrating the intensities of $\Lambda(R,\theta)$ with respect to $\theta$ at different radius $R$ which is measured from the inferred center of the quarter circle used to fit the $\vec{q}_4$ trajectory. Center of the quarter circle positions are doping dependent moving on the $\vec{q} = (0,0) - (2\pi,2\pi)$ line; $\theta$ is the angle measured from the $\vec{q} = (0,0) - (2\pi,2\pi)$ line.

$$W(R) = \int_{\theta_0}^{\theta_1} \Lambda(R,\theta)d\theta \tag{S8}$$

where $\theta_0 \approx 1/4 \times \pi/4$, $\theta_1 \approx 1.3 \times \pi/4$ for $p < 0.19$ and $\theta_1 = \pi/4$ for $p > 0.19$. $\theta_0$, that is typically $\pi/16$, is set to avoid irrelevant signal which is generated by other *q*-vectors near $\vec{k} = (\pi/2,\pi/2)$. $W(R)$ is a smooth function with a single broad radial peak encapsulating $R_{\vec{q}_4}$ where the $\vec{q}_4$ trajectory exists, and decreases as $|R - R_{\vec{q}_4}|$ increases. Then, $W(R)$ is multiplied by the original $\Lambda(R,\theta)$ to enhance the experimentally determined $\vec{q}_4$ trajectory while suppressing the unnecessary signals. Fig. S3 shows the doping dependence of the resulting normal state Fermi surface. Dots are the location of the Fermi surface. Note that signals near the nodal regions are ambiguous since $\vec{q}_4$ signals are not clear in the underdoped regime and are mixed with other *q*-vectors in the overdoped regime. Since $W(R)$ is obtained by the assumption that the $\vec{q}_4$ trajectory is circular, applying this "filter" may alter the $\vec{q}_4$ trajectory slightly. However, as shown in Fig. S3, results are consistent with those shown in Fig. S2 and key features are vividly reproduced at the qualitative level - the Fermi surface topology change from "arc" to closed and hole-like contour centered at $(\pi,\pi)$ occurs at $p\approx0.19$. This can be seen very clearly and directly in Movie S2.

### (III) Doping dependence of Q≠0 incommensurate modulations

In order to analyze the spatial variation of the broken symmetry in $Z(\vec{r}, e = 1)$ as a function of doping, first, energy scale is normalized by each local $\Delta_1$, namely, $e \equiv E/\Delta_1(\vec{r})$, so that all the electronic broken symmetries in $Z(\vec{r}, E)$ appears at $e$=1.[35] Second, picometer scale distortion of the image from imperfect scanner effects and the crystal local distortion is corrected by Lawler-Fujita algorithm described in the ref [11, 35]. The results are electronic structure images further transformed into square lattice with perfect global phase coherence in order to characterize the local electronic symmetry breaking with the highest precision. Finally, Fourier transform of $Z(\vec{r}, e = 1)$ is taken to visualize the $q$-space electronic structure $Z(\vec{q}, e = 1)$. In Fig. S4 and S5, $Z(\vec{r}, e = 1)$ and $Z(\vec{q}, e = 1)$ respectively at different doping ranging from strongly underdoped ($p \approx 0.06$) to strongly overdoped ($p \approx 0.23$) regime are shown. In the strongly underdoped regime, the intra-unit-cell $C_4$ symmetry breaking and the incommensurate modulations that break both rotational and translational symmetry are clearly seen.

It has long been known that the dispersive Bogoliubov QPI vector $\vec{q}_5$ apparently converts to quasistatic wavevector $\vec{q}_5^*$ (or $S$) above QPI extinction energy[32]. Thus, when the incommensurate modulation $\vec{q}_5^*$ is analyzed, one has to distinguish $\vec{q}_5^*$ from $\vec{q}_5$ whose magnitudes are very similar. To extract $\vec{q}_5^*$ attenuating the $\vec{q}_5$ modulations, we took sum of $Z(\vec{r}, e)$ with respect to $e$ and then divided by the number of layers for the normalization. In this procedure, dispersive signature disappears after the sum since the spatial phase of the modulations is nearly random canceling the intensities of dispersive signature, while $\vec{q}_5^*$ (or $S$) is enhanced due to the non-dispersive characteristics. Then, we extract the intensities of $S$ ($I(S)$) at different dopings and plotted in Fig. 3G in the main text. Line profiles along (1,0) and (0,1) in $Z(\vec{q}, e)$ are fitted by Lorentzian and $I(S)$ is determined as an obtained amplitude averaged for both (1,0) and (0,1) directions. We have also performed two other independent methods in determining the $I(S)$ and all showed consistent results.

### (IV) Doping dependence of $O_N^q(\vec{r}, e = 1)$

Before showing the doping dependence of $O_N^q(\vec{r}, e = 1)$, we show in Fig. S6 an example of $C_4$ symmetry breaking detected in $q$-space. Fig. S6A is a typical Re[$Z(\vec{q}, e = 1)$] from underdoped sample (UD45K). Re[$Z(\vec{q}, e = 1)$] represents two major incommensurate and commensurate electronic modulations. Black arrows indicate the commensurate peaks for $x$ and $y$ direction, and they are compared in Fig. S6B exhibiting the inequivalent peak height between *x* and *y* direction that indicates the electronic $C_4$ symmetry breaking for commensurate modulation in $Z(\vec{r}, e=1)$.

$C_4$ symmetry breaking at given location, $O_N^q(\vec{r})$ is defined by

$$O_N^q(\vec{r}) = \frac{\text{Re}[Z(\vec{r}, \vec{Q}_y)] - \text{Re}[Z(\vec{r}, \vec{Q}_x)]}{\text{Re}[Z(\vec{r}, \vec{Q}_y)] + \text{Re}[Z(\vec{r}, \vec{Q}_x)]} \tag{S9}$$

where $\vec{Q}_x$ and $\vec{Q}_y$ are the Bragg vectors along two orthogonal Cu-O directions. $Z(\vec{r}, \vec{Q})$ is a Fourier filtered $Z(\vec{r})$ with respect to $\vec{Q}$ and is obtained by

$$Z(\vec{r}, \vec{Q}, e) = \sum_{\vec{R}} Z(\vec{R}, e) e^{i\vec{Q}\cdot\vec{R}} e^{-\frac{|\vec{R}-\vec{r}|^2}{2\Gamma^2}} \frac{1}{2\pi\Gamma^2} \tag{S10}$$

where $\Gamma$ is the cut-off length for the spatial average which is set as 25Å and is indicated by the broken circle in Figure 3F and S7F.

Fig. S7 shows a doping dependence of $O_N^q(\vec{r}, e=1)$. In the strongly underdoped regime, this oxygen-site-specific measure of $C_4$ symmetry is strongly broken toward one direction all over the FOV. However, with increasing doping, spatial variation of the $C_4$ symmetry breaking becomes disordered exhibiting the increase of the spatial fluctuation of the $C_4$ symmetry breaking.

Quantitatively, doping dependence of the $O_N^q(\vec{r})$ is characterized by estimating a spatial average and fluctuation, which are respectively defined as below,

$$\langle O_N^q \rangle = \frac{1}{n} \sum_{\vec{r}} O_N^q(\vec{r})$$

$$\delta O_N^q = \sqrt{\frac{1}{n-1} \sum_{\vec{r}} \left(O_N^q(\vec{r}) - \langle O_N^q \rangle\right)^2} \tag{S11}$$

where *n* is the number of pixels. The spatial average $\langle O_N^q \rangle$ estimates the global $C_4$ symmetry breaking in the image, while the rms average $\delta O_N^q$ is the measure of spatial fluctuation in $O_N^q(\vec{r})$. Fig. S8 shows the doping dependence of the $O_N^q(\vec{r})$ histogram. As plotted in Fig. 3H in the main text, $\langle O_N^q \rangle$ decreases with increasing doping and almost goes to zero near *p*~0.19, while $\delta O_N^q$ (width of the histogram) increases with increasing doping and $\delta O_N^q$ becomes significantly large near *p*~0.19.

**(V) Challenges in measuring intra-unit-cell nematicity**

In an interesting recent study[43] some possible procedures for measurement of nematicity in STM experiments were examined. Although the conclusions of that work, as they apply to the data presented in Ref. [43] may be true, they have no relationship to or implications for those presented in this paper or in Ref. [11, 26, 35, 36]. The reason is because Ref. [43] defined electronic "nematicity" to be the ratio of the magnitudes measured at the $\vec{Q}_x = [10]$ and $\vec{Q}_y = [01]$ Bragg peaks of an image $g(\vec{r})$ :

$$\mathcal{N} = \frac{|\tilde{g}(\vec{Q}_y)| - |\tilde{g}(\vec{Q}_x)|}{|\tilde{g}(\vec{Q}_y)| + |\tilde{g}(\vec{Q}_x)|}, \tag{S12}$$

This is both mathematically and physically fundamentally different from the definition of oxygen-site-specific intra-unit-cell nematicity used in our studies:

$$O_N^q \propto \left[ Re[\tilde{g}(\vec{Q}_y)] - Re[\tilde{g}(\vec{Q}_x)] \right]. \tag{S13}$$

The mathematical distinction is that $O_N^q$ is a lattice-phase sensitive measure and thus an atomic site selective order parameter. It is characterizing the contributions primarily from the oxygen atoms in the CuO$_2$ plane[11, 35]. By contrast Eqn. S12 mixes both real and imaginary Fourier Bragg components promiscuously and cannot achieve the crucial oxygen site selectivity. This is a key point because we have previously demonstrated using independent *r*-space visualization of intra-unit-cell symmetry[11, 35, 36] that $O_N^q$ is actually derived specifically from electronic differences between the two oxygen sites within each CuO$_2$ unit cell.

In order to obtain this oxygen-site selective measure of electron nematicity $O_N^q$, which is at the core of our studies[11, 26, 35, 36], and is given by Eqn. S13, one has first to correct for ~picometer per unit cell scale distortions in rectilinearity of the images of $g(\vec{r},E)$, and then establish the lattice phase with high precision (2% of 2π). We have introduced a mathematical technique for doing so that is described in complete details in Hamidian *et al*.[44] This lattice-phase-definition technique generates a square lattice with perfect global phase coherence with an atomic peak at the Cu/Bi origin; this is equivalent to ensuring that there are no imaginary (sine) components to the Bragg peaks in the Fourier transform of the topograph $T(\vec{r})$. In this procedure one undoes distortions in the raw $T(\vec{r})$ data – transforming into a distortion-corrected topograph $T'(\vec{r})$ exhibiting the known periodicity and lattice-phase-definition. To demonstrate success in this procedure, one must show that, in the Fourier transform of corrected topograph $T'(\vec{r})$, the $Re[T'(\vec{q})]$ Bragg peaks are single pixels and $Im[T'(\vec{q})]$

Bragg peaks are nearly zero. The identical geometrical transformations yielding $T'(\vec{r})$, are also carried out on every $g(\vec{r},E)$ acquired simultaneously with the $T(\vec{r})$ to yield a distortion corrected $g'(\vec{r},E)$. The $T'(\vec{r})$ and $g'(\vec{r},E)$ are then registered to each other and to the lattice with excellent periodicity and lattice-phase-definition. Only after the above lattice-phase-definition procedures are demonstrated successfully (e.g., Fig. S9) can the utility of $O_N^q$ in Eqn. S13 for detecting oxygen-site-specific intra-unit-cell nematicity be evaluated meaningfully. However, these lattice-phase-definition procedures were not demonstrated or evaluated in Ref. [43]; therefore no deductions about the validity of Eqn. S13 to evaluate oxygen-site-specific intra-unit-cell $C_{4v}$ symmetry breaking in our [11, 26, 35, 36] STM data should be made therefrom.

Physically the two approaches in S12 and S13 are also completely different. This is because, at a practical level, $O_N^q$ in Eqn. S13 not only requires the above mathematical steps to be carried out successfully, but it also requires spectroscopic measurements using approximately 50 pixels inside each $CuO_2$ unit-cell in a large enough FOV so that a $q$-space resolution approaches 1% of $2\pi/a$. If these measurement specifications are not achieved demonstrably, no deductions about the validity of Eqn. S13 to evaluate oxygen-site-specific intra-unit-cell $C_{4v}$ symmetry breaking in our STM data[11, 26, 35, 36] can or should be made.

Conversely, there is a simple practical test that reliably demonstrates the capability to correctly detect intra-unit-cell rotational symmetry breaking using STM data with Eqn. S13. If nearby domains (or even unit cells[11, 36]) of opposite nematic order are observed using same tip, then tip anisotropy is obviously immediately ruled out as the cause of rotational symmetry breaking in the data. Clearly Fig. 3F of the main text and Fig. S7 show just such domains, demonstrating that any putative anisotropy in the tip is not the cause of the oxygen-site-specific intra-unit-cell symmetry breaking measured and reported herein or in [11, 26, 35, 36].

Therefore in sum, because:

1. Eqn. S12, the measure of nematicity studied in Ref. [43], is mathematically and physically distinct from the measure ($O_N^q$ in Eqn. S13) used in our research;

2. The lattice-phase-definition procedures that are indispensable to oxygen-site-specific intra-unit-cell nematicity $O_N^q$ were not applied or evaluated in Ref. [43];

3. The repeated direct atomic-scale demonstration, both in our studies[11, 26, 35, 36] and that of other groups[45] that tip anisotropy does not influence the capability to distinguish breaking of rotational symmetry between the two oxygen atoms in each $CuO_2$ unit cell;

one must conclude that the issues pointed out in Ref. [43] in measuring intra-unit-cell rotational symmetry breaking do not pertain to any of our studies, including Refs. [11, 26, 35, 36] and the present work.

**Captions for Figures**

**Figure S1**. (A1-E1) Model spectral function $A(\vec{k}, \omega)$ for the $d$-wave superconductor at different energies. (A2-E2) $\Lambda(\vec{q})$ summed up to the given energy visualizing the evolution of the normal state Fermi surface. Here, we used $V_S = 0.1$ eV and $V_m = 0$ eV in the calculation plus the band and gap parameters in the text.

**Figure S2**. Doping dependence of $\Lambda(\vec{q})$ visualizing the normal state Fermi surface obtained by $\vec{q}_4$, and showing the abrupt change in topology of Fermi surface at critical doping $p_c \approx 0.19$.

**Figure S3**. Doping dependence of the filtered Fermi surface location as visualized by $\vec{q}_4$.

**Figure S4**. Real space evolution of the broken symmetry in $Z(\vec{r}, e = 1)$ at different doping from underdoped to overdoped regime across the critical doping $p_c \approx 0.19$.

**Figure S5**. $q$-space representation of the broken symmetry in $Z(\vec{r}, e = 1)$ at different doping from underdoped to overdoped regime across the critical doping at $p_c \approx 0.19$. The peak associated with incommensurate modulation decreases with increasing doping and become virtually undetectable at $p$=0.19.

**Figure S6**. A. $\text{Re}[Z(\vec{q}, e = 1)]$ B. Line profile along $x$ and $y$ direction in A exhibiting an inequivalent Bragg peak intensities.

**Figure S7**. Doping dependence of $O_N^q(\vec{r}, e = 1)$.

**Figure S8**. Doping dependence of the histogram of all $O_N(\vec{r})$ values in each field of view. This shows that nematicity is due to an imbalance in the number of unit cells of one Ising nematic value, compared to the other - as originally explained in Ref. [11].

**Figure S9**. Raw topograph $T(\vec{r})$(A), and raw differential conductance $g(\vec{r})$(B) with its Fourier transform(C) are compared with their distortion-corrected maps in (D), (E), and (F). Proper application

of the distortion-correction to achieve the necessary lattice-phase-definition is ensured by the resulting Bragg peaks width in (D) and (F) as shown in the insets.

**Captions for Movies**

**Movie S1.** Typical energy dependence of $q$-space electronic structure $Z(\vec{q}, E)$. At low energies dispersive peaks associated with the interference of the Bogoliubov quasiparticles are seen and these intensities decrease with increasing energy, and then replaced with ultra-slow dispersive peaks.

**Movie S2.** Evolution of the $k$-space locus of coherent Bogoliubons - which is ascribed to the normal state Fermi surface - as a function of hole density. Fermi surface topology rapidly changes near the critical doping $p \approx 0.19$ from the arc to the hole-like Fermi surface centered at $\vec{k} = (\pi, \pi)$.

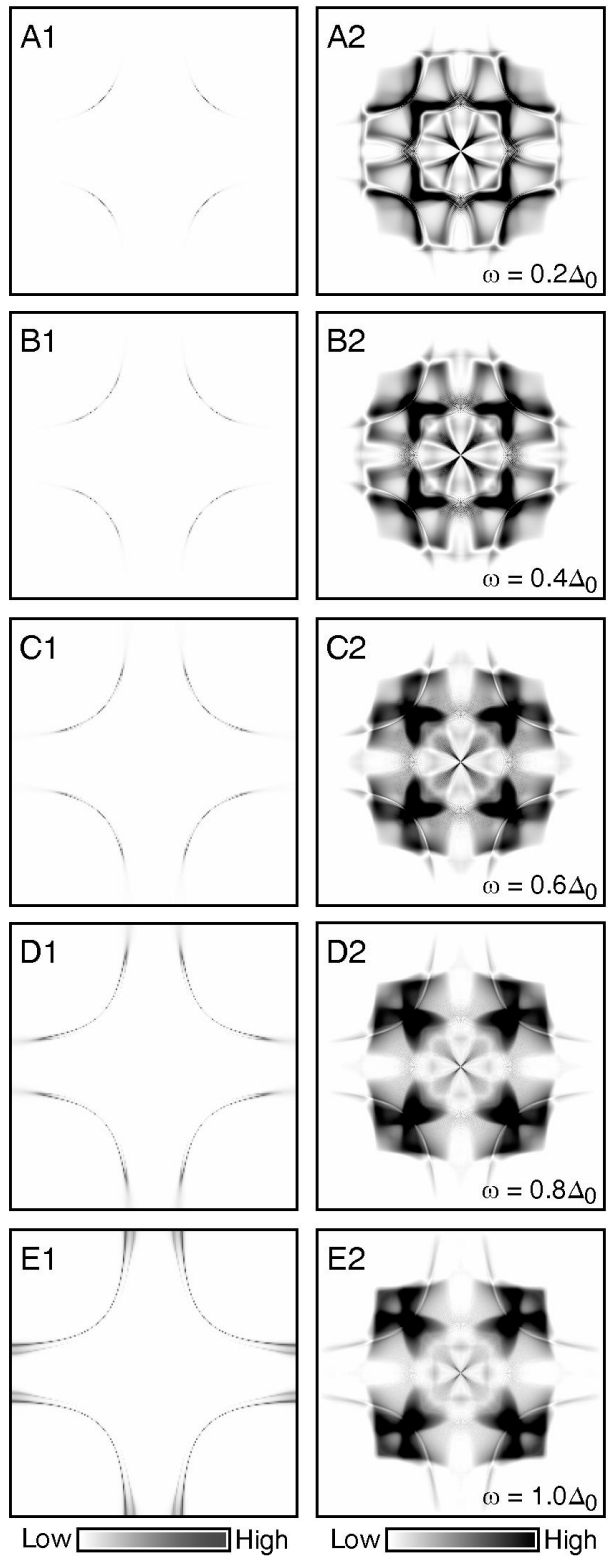

Fig. S1

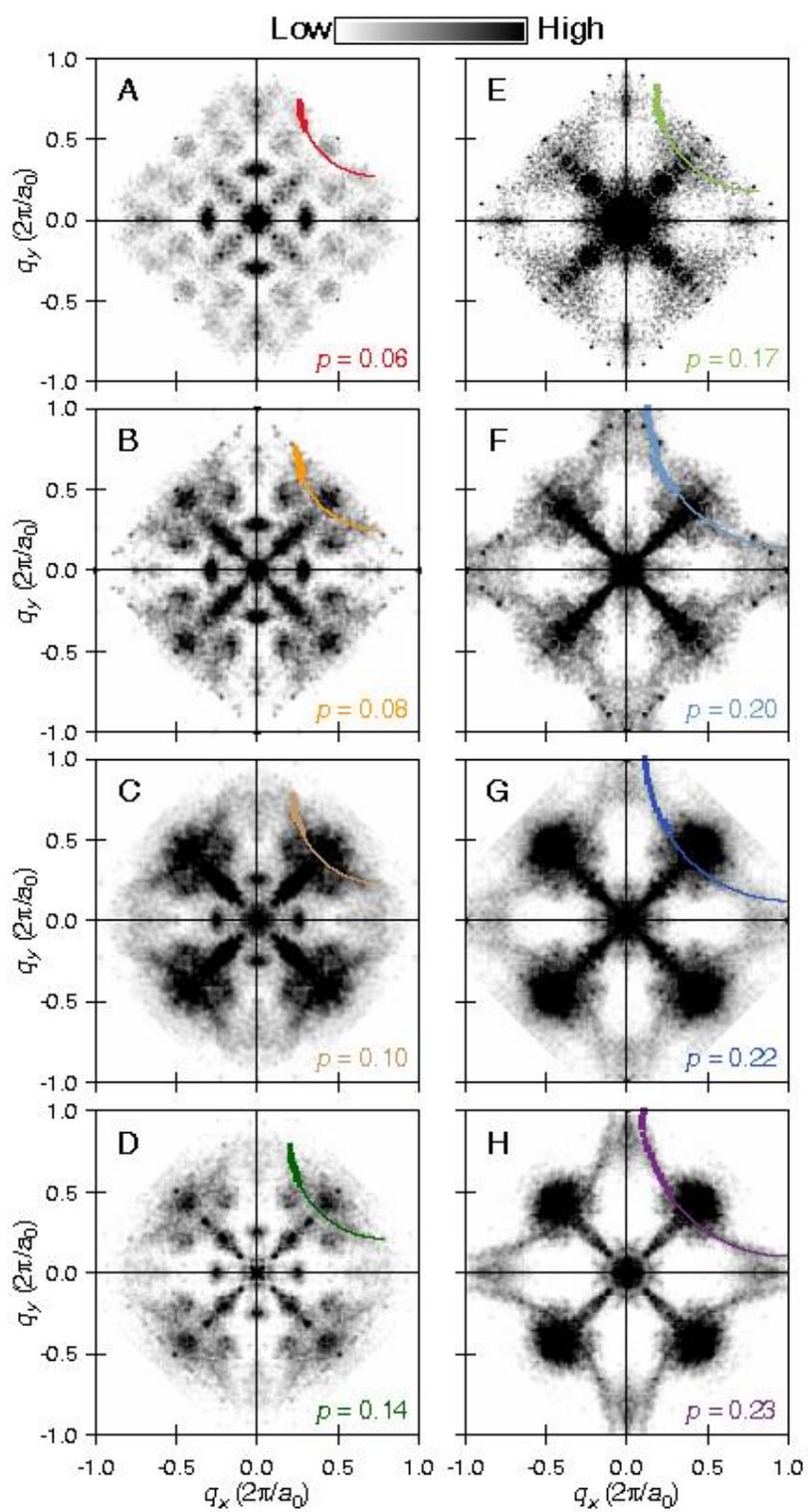

Fig. S2

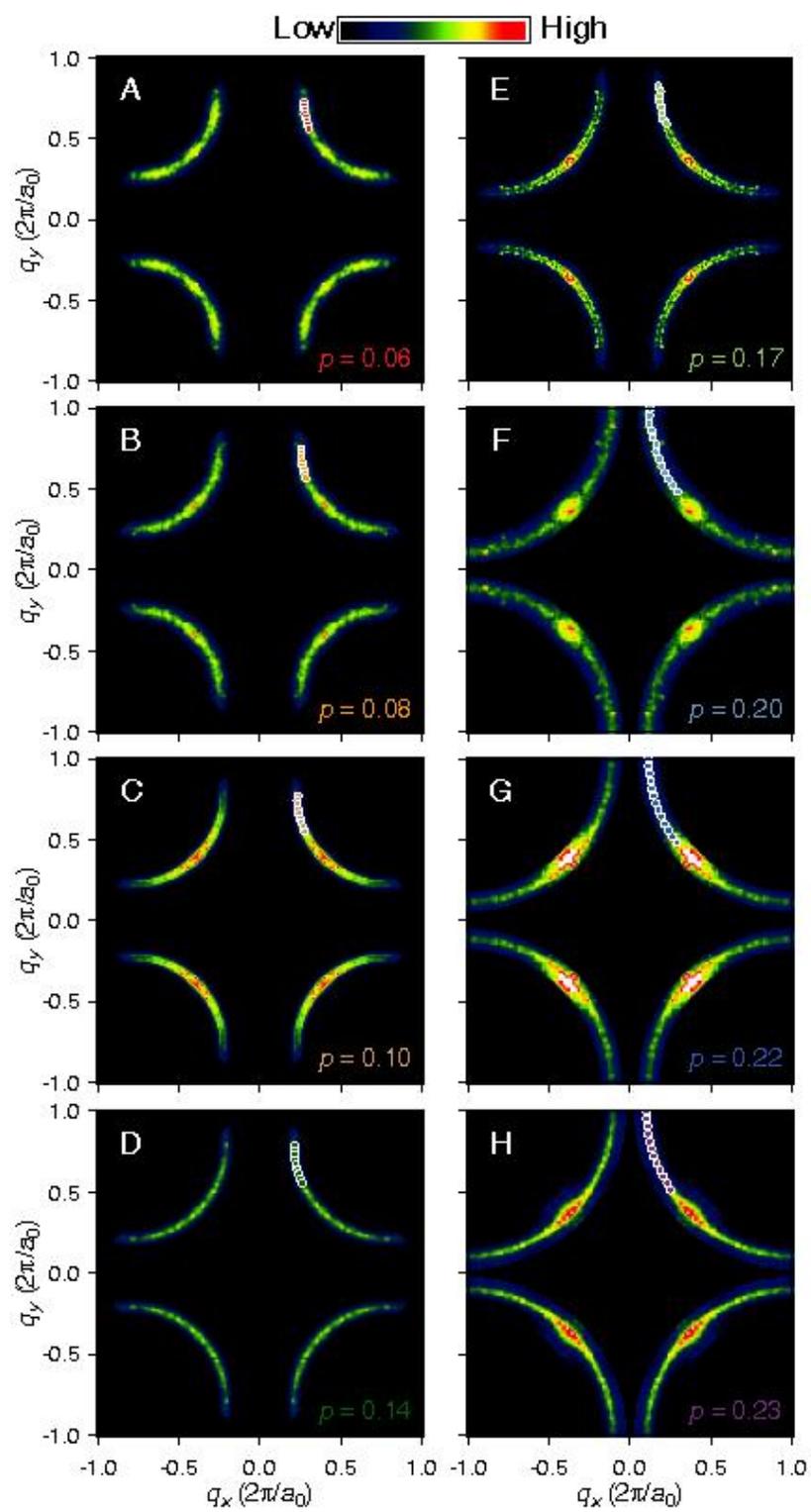

Fig. S3

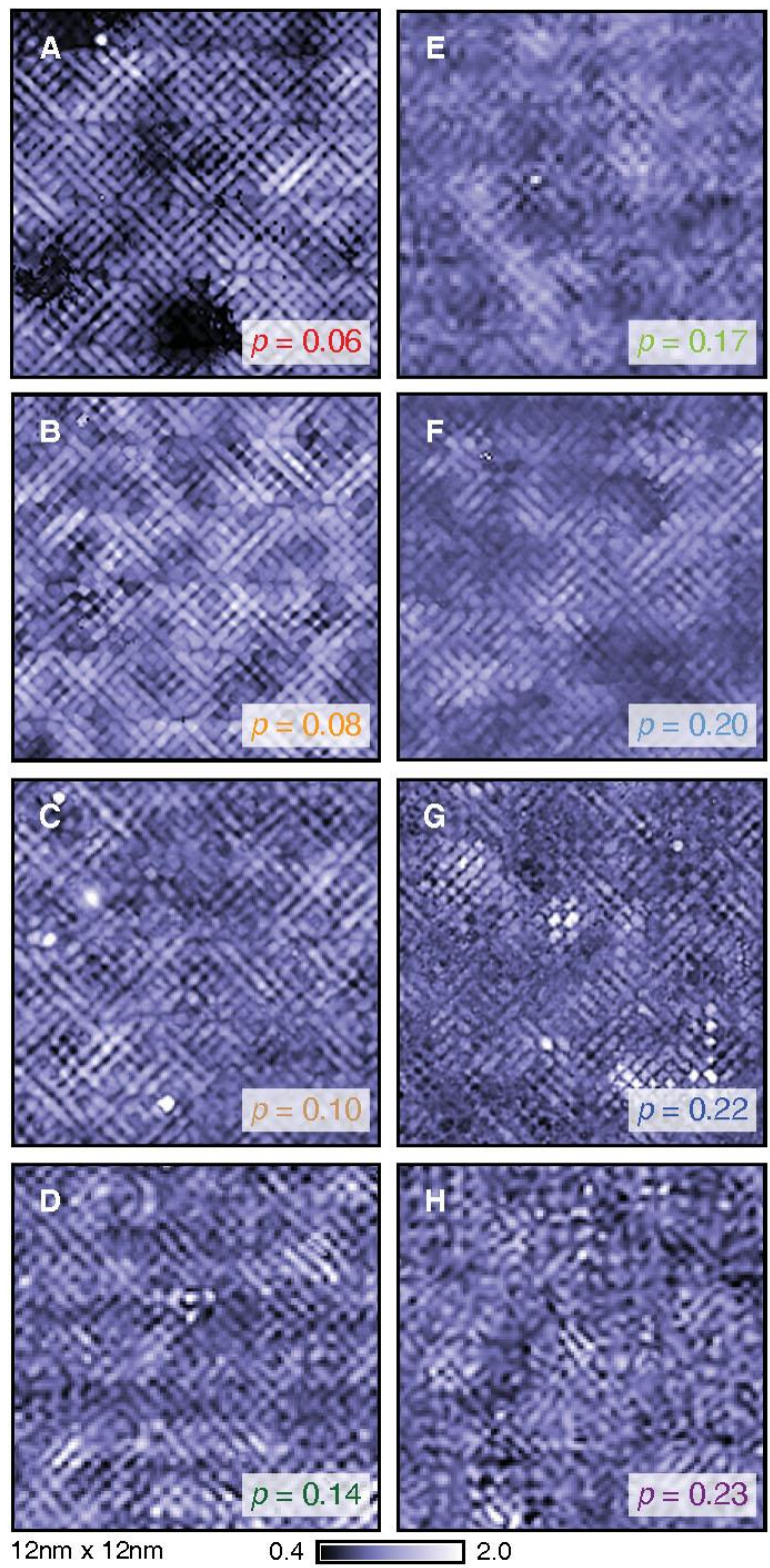

Fig. S4

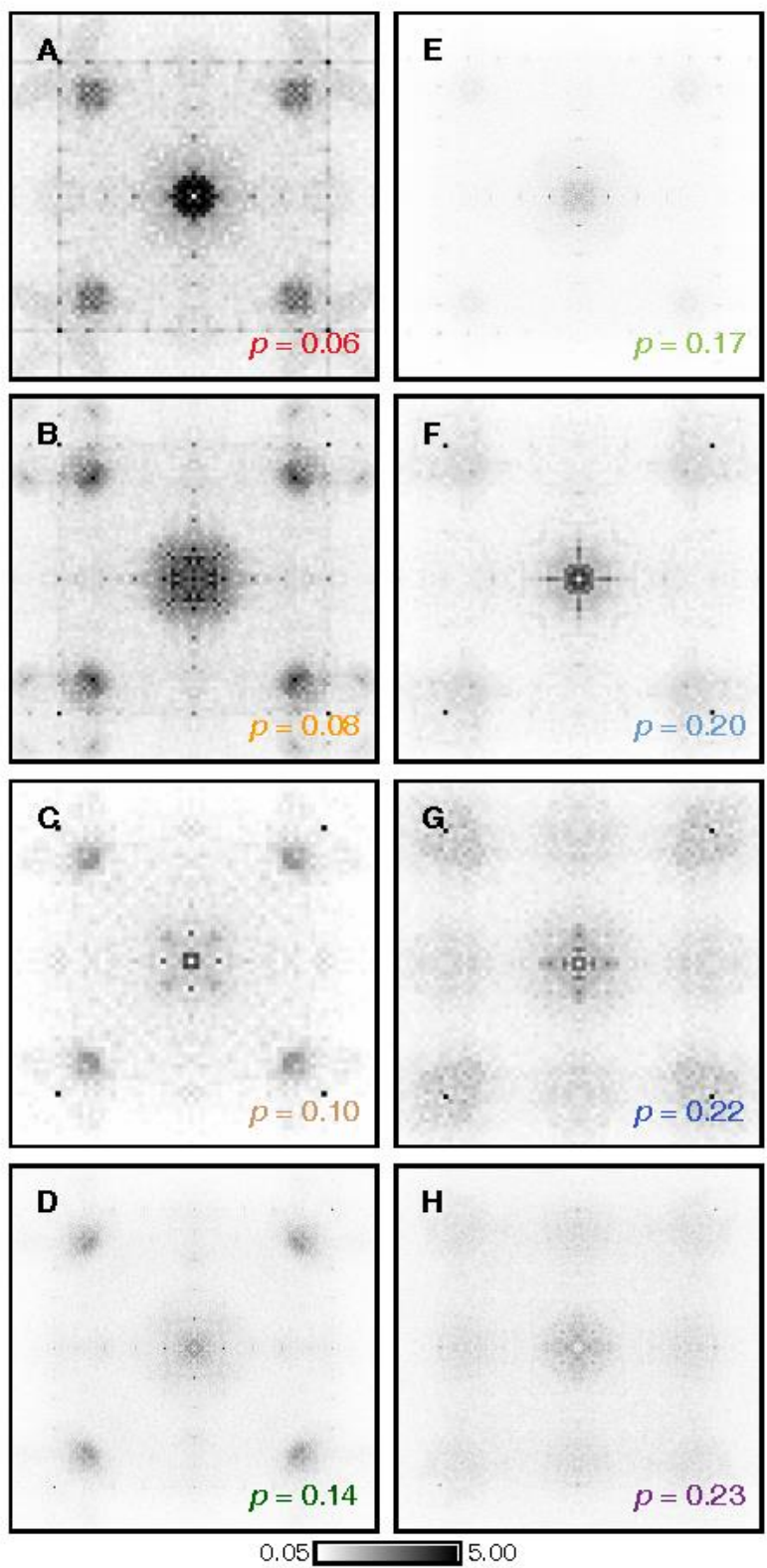

Fig. S5

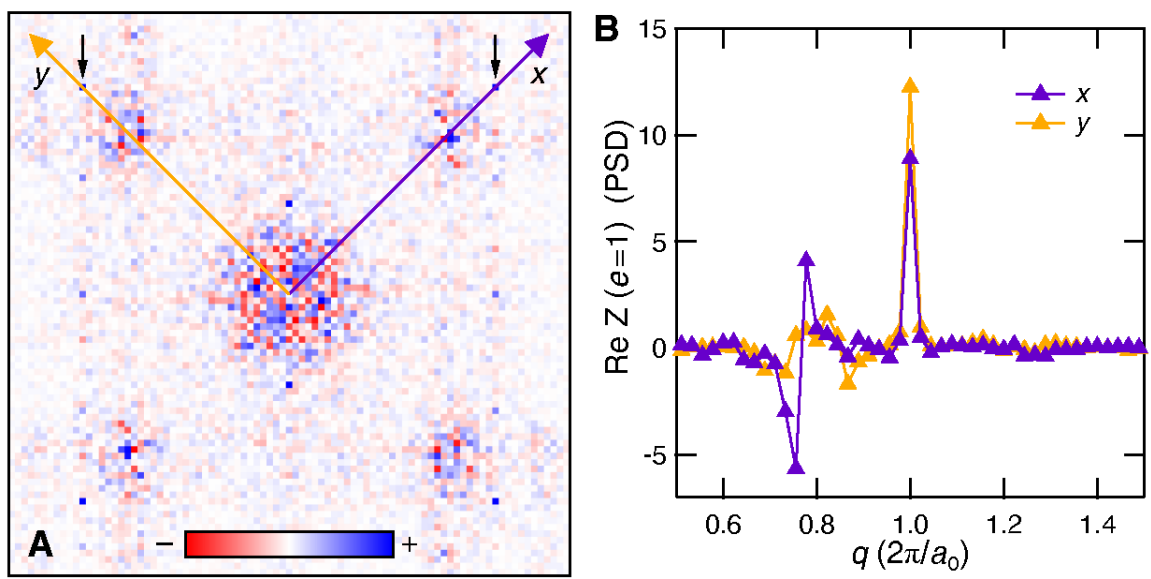

Fig. S6

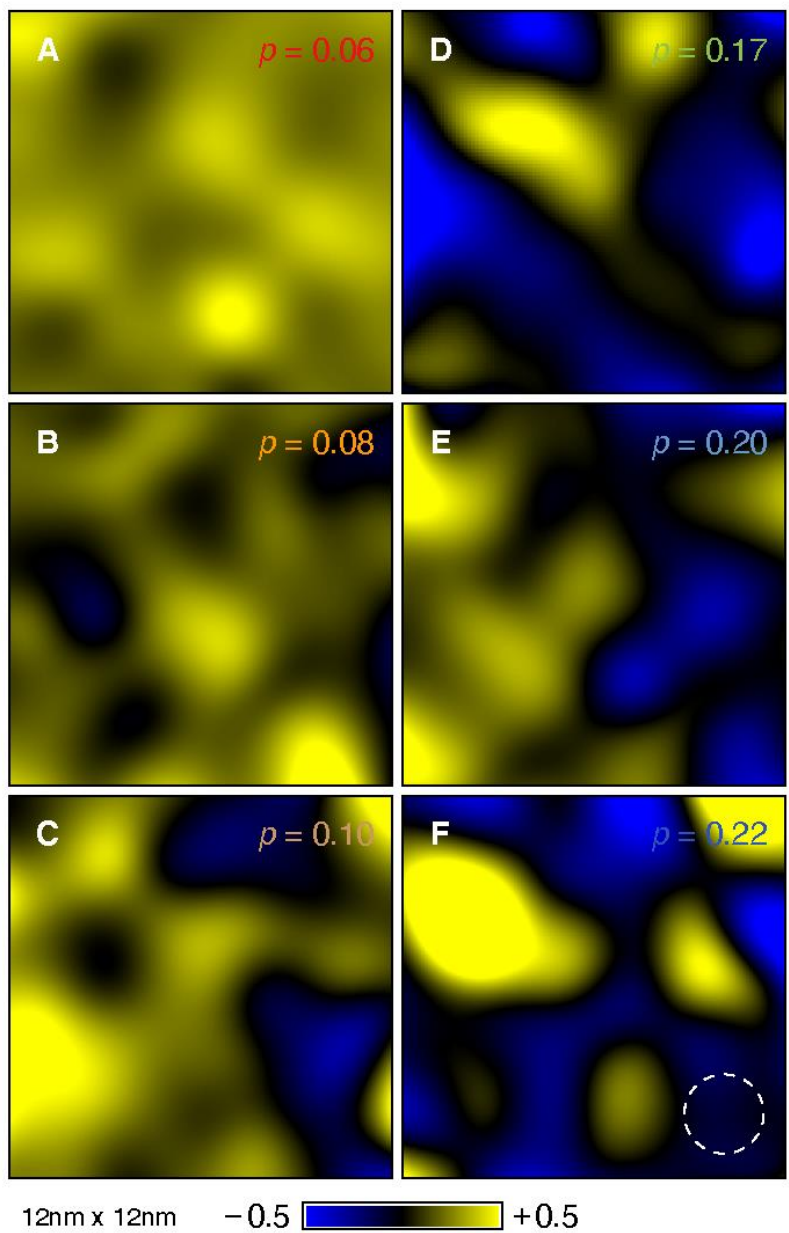

Fig. S7

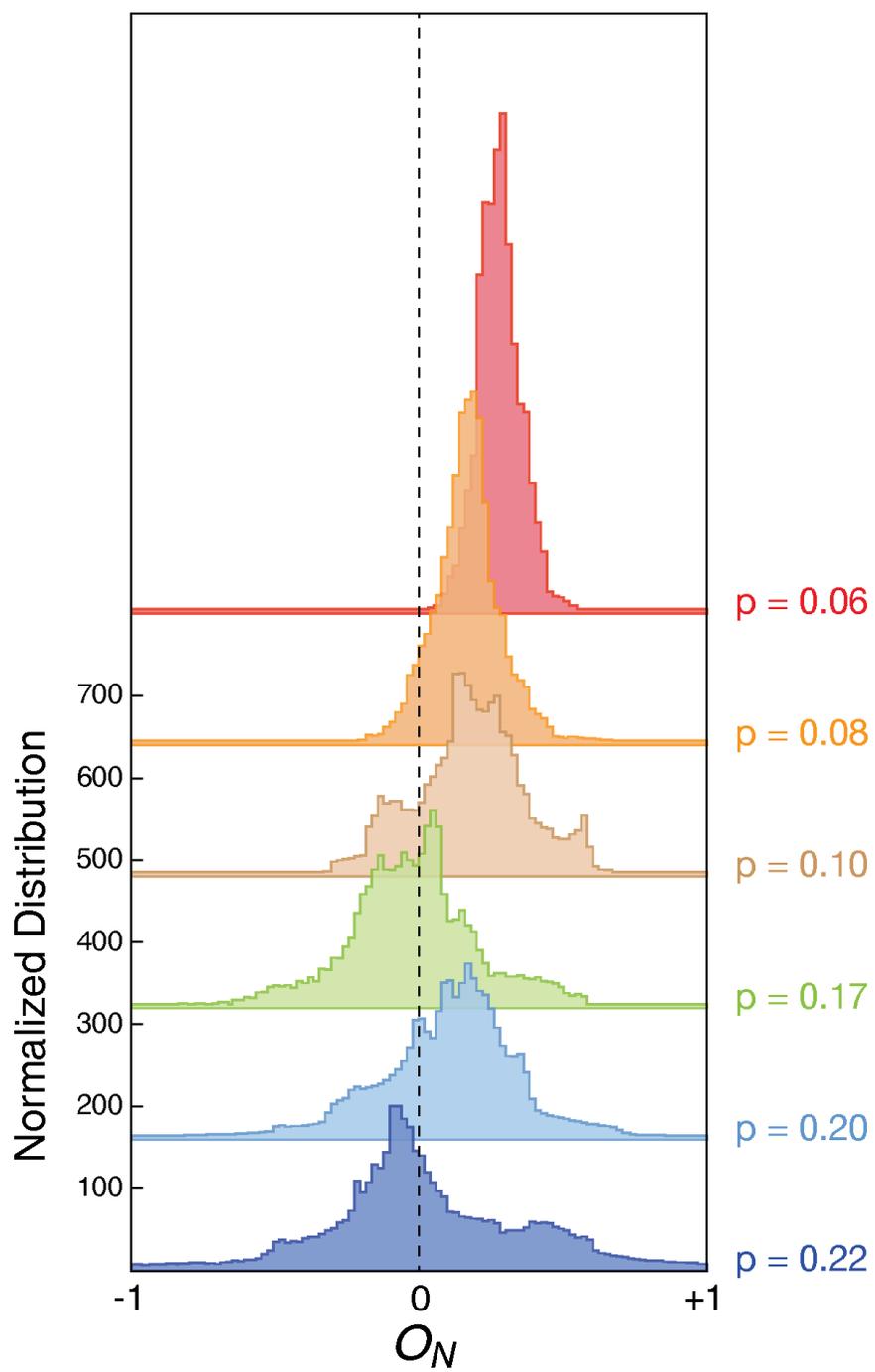

Fig. S8

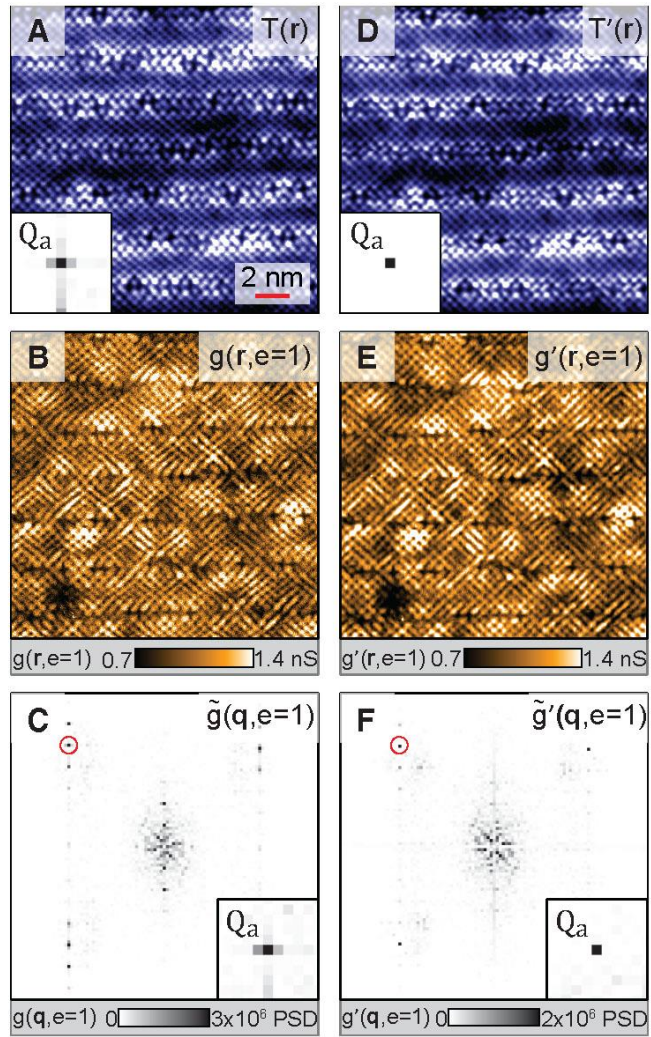

Fig. S9